\documentstyle[aps
,epsfig%
,preprint%
]{revtex}
\begin{document}
\draft
\preprint{Guchi-TP-002}
\date{\today
}
\title{
Boson Stars with Large Self-interaction in (2+1) dimensions: 
an Exact Solution
}
\author{Kenji~Sakamoto${}^1$ and
Kiyoshi~Shiraishi${}^{1,2}$%
\thanks{e-mail: {\tt g00345@simail.ne.jp,
shiraish@sci.yamaguchi-u.ac.jp}
}}
\address{${}^1$Graduate School of Science and Engineering, 
Yamaguchi University\\ 
Yoshida, Yamaguchi-shi, Yamaguchi 753-8512, Japan
}
\address{${}^2$Faculty of Science, Yamaguchi University\\
Yoshida, Yamaguchi-shi, Yamaguchi 753-8512, Japan
}
\maketitle
\begin{abstract}
An exact solution for a nonrotating boson star in $(2+1)$ dimensional
gravity with a negative cosmological constant is found.
The relations among mass, particle number, and radius of
the $(2+1)$ dimensional boson star are shown.
\end{abstract}




Self-gravitating systems have been investigated in various situations.
Boson stars (\cite{BS}\cite{LM} for reviews) have a very simple
constituent,  a complex scalar field which is bound by gravitational
attraction. Thus the boson star provides us with the simplest model of
relativistic stars.

The solutions for relativistic boson stars are only numerically 
obtained in four dimensions. In $(2+1)$ dimensions, static equilibrium
configurations have been argued \cite{CZ} in Einstein gravity with a
negative cosmological constant.

In this paper, we obtain an exact solution for a nonrotating boson 
star in $(2+1)$ dimensional gravity with a negative cosmological 
constant. We consider that the scalar field has a strong 
self-interaction.
An infinitely large self-interaction term in the model leads to much
simplications as in the $(3+1)$ dimensional case~\cite{CSW}.
The study of exact solutions will lead to a new aspect of gravitating
systems and clarify the similarity and/or the difference among the 
other dimensional cases.


We consider a complex scalar field with mass $m$ and a quartic 
self-coupling constant $\lambda$.
The action for the scalar field coupled to gravity can be
written down as
\begin{equation}
S=\int d^3x\sqrt{-g}\left[
\frac{1}{16\pi G}\left(R+2C\right)-
\left|\nabla_{\mu}\varphi\right|^2-
m^2\left|\varphi\right|^2-
\frac{\lambda}{2}\left|\varphi\right|^4\right],
\label{eq:action}
\end{equation}
where $R$ is the scalar curvature and the positive constant
$C$ stands for the (negative) cosmological constant.
$G$ is the Newton constant.

Varying the action (\ref{eq:action}) with respect to
the scalar field and the metric yields equations of
motion. The equation of motion for the scalar field is
\begin{equation}
\nabla^2\varphi-m^2\varphi-\lambda\left|\varphi\right|^2\varphi=0,
\label{eq:em}
\end{equation}
while the Einstein equation is
\begin{equation}
R_{\mu\nu}-\frac{1}{2}g_{\mu\nu}R=
8\pi G\left[
2{\rm Re}\left(\nabla_{\mu}\varphi^{*}\nabla_{\nu}\varphi-
\frac{1}{2}\left|\nabla_{}\varphi\right|^2 g_{\mu\nu}\right)-
m^2\left|\varphi\right|^2g_{\mu\nu}-\frac{\lambda}{2}
\left|\varphi\right|^4g_{\mu\nu}\right]+Cg_{\mu\nu}.
\label{eq:Ee}
\end{equation}

We assume the three-dimensional metric 
for a static circularly symmetric spacetime as
\begin{equation}
ds^2=-\alpha^2(r)dt^2+\beta^2(r)dr^2+
\frac{\gamma^2(r)}{\gamma_0^2}d\theta^2,
\label{eq:met}
\end{equation}
where $\alpha$, $\beta$, and $\gamma$ are functions of the radial
coordinate $r$ only, while $\gamma_0$ is a constant.
Because of the coordinate invariance with respect to $r$, we have 
a freedom in choosing the function $\beta(r)$.
We will use this residual ``gauge choice'' later.

We also assume that the complex scalar has 
a phase which is linear in the temporal coordinate:
\begin{equation}
\varphi=e^{-i\omega t}\varphi(r),
\end{equation}
where $\varphi(r)$ is a function of the radial coordinate $r$ only 
and $\omega$ is a constant.

Then the equations~(\ref{eq:em}) and (\ref{eq:Ee}) are written as
\begin{eqnarray}
\frac{\omega^2}{\alpha^2}\varphi&+&\frac{1}{\alpha\beta\gamma}
\frac{d}{dr}
\left(\frac{\alpha\gamma}{\beta}\frac{d\varphi}{dr}\right)-
m^2\varphi-\lambda\varphi^3=0, \\
\frac{1}{\beta\gamma}\frac{d}{dr}
\left(\frac{1}{\beta}\frac{d\gamma}{dr}\right)&=&
8\pi G\left[
-\frac{1}{\beta^2}\left(\frac{d\varphi}{dr}\right)^2-
\frac{\omega^2}{\alpha^2}\varphi^2-m^2\varphi^2-
\frac{\lambda}{2}\varphi^4\right]+C, \\
\frac{1}{\alpha\beta^2\gamma}\frac{d\alpha}{dr}
\frac{d\gamma}{dr}&=&
8\pi G\left[
+\frac{1}{\beta^2}\left(\frac{d\varphi}{dr}\right)^2+
\frac{\omega^2}{\alpha^2}\varphi^2-m^2\varphi^2-
\frac{\lambda}{2}\varphi^4\right]+C, \\
\frac{1}{\alpha\beta}\frac{d}{dr}
\left(\frac{1}{\beta}\frac{d\alpha}{dr}\right)&=&
8\pi G\left[
-\frac{1}{\beta^2}\left(\frac{d\varphi}{dr}\right)^2+
\frac{\omega^2}{\alpha^2}\varphi^2-m^2\varphi^2-
\frac{\lambda}{2}\varphi^4\right]+C.
\end{eqnarray}

For convenience, we rescale the variables as
\begin{equation}
\tilde{r}=mr,~~~\tilde{\varphi}=\sqrt{8\pi G}\varphi,~~~
\Lambda=\frac{\lambda}{8\pi Gm^2},~~~
\tilde{\omega}=\omega/m,~~~and~~~\tilde{C}=C/m^2.
\end{equation}
Using these variables, we can rewrite the equations as
\begin{eqnarray}
\frac{\tilde{\omega}^2}{\alpha^2}\tilde{\varphi}&+&
\frac{1}{\alpha\beta\gamma}
\frac{d}{d\tilde{r}}
\left(\frac{\alpha\gamma}{\beta}
\frac{d\tilde{\varphi}}{d\tilde{r}}\right)-
\tilde{\varphi}-\Lambda\tilde{\varphi}^3=0, \\
\frac{1}{\beta\gamma}\frac{d}{d\tilde{r}}
\left(\frac{1}{\beta}\frac{d\gamma}{d\tilde{r}}\right)&=&
-\frac{1}{\beta^2}\left(\frac{d\tilde{\varphi}}{d\tilde{r}}\right)^2-
\frac{\tilde{\omega}^2}{\alpha^2}\tilde{\varphi}^2-\tilde{\varphi}^2-
\frac{\Lambda}{2}\tilde{\varphi}^4+\tilde{C}, \\
\frac{1}{\alpha\beta^2\gamma}\frac{d\alpha}{d\tilde{r}}
\frac{d\gamma}{d\tilde{r}}&=&
+\frac{1}{\beta^2}\left(\frac{d\tilde{\varphi}}{d\tilde{r}}\right)^2+
\frac{\tilde{\omega}^2}{\alpha^2}\tilde{\varphi}^2-\tilde{\varphi}^2-
\frac{\Lambda}{2}\tilde{\varphi}^4+\tilde{C}, \\
\frac{1}{\alpha\beta}\frac{d}{d\tilde{r}}
\left(\frac{1}{\beta}\frac{d\alpha}{d\tilde{r}}\right)&=&
-\frac{1}{\beta^2}\left(\frac{d\tilde{\varphi}}{d\tilde{r}}\right)^2+
\frac{\tilde{\omega}^2}{\alpha^2}\tilde{\varphi}^2-\tilde{\varphi}^2-
\frac{\Lambda}{2}\tilde{\varphi}^4+\tilde{C}.
\end{eqnarray}


Next, we rescale the variables again, to study the limit of
large self-inteaction.
New variables are:
\begin{equation}
r_{*}=\tilde{r}/\sqrt{\Lambda},~~~
\varphi_{*}=\sqrt{\Lambda}\tilde{\varphi},~~~
and~~~C_{*}=\Lambda\tilde{C}.
\end{equation}
Then the equations can be written as
\begin{eqnarray}
\frac{\tilde{\omega}^2}{\alpha^2}\varphi_{*}&+&
\frac{1}{\Lambda}\frac{1}{\alpha\beta\gamma}
\left(\frac{\alpha\gamma}{\beta}\varphi_{*}'\right)'-
\varphi_{*}-\varphi_{*}^3=0, \\
\frac{1}{\beta\gamma}\left(\frac{\gamma'}{\beta}\right)'&=&
-\frac{1}{\Lambda}\frac{1}{\beta^2}\left(\varphi_{*}'\right)^2-
\frac{\tilde{\omega}^2}{\alpha^2}\varphi_{*}^2-\varphi_{*}^2-
\frac{1}{2}\varphi_{*}^4+C_{*}, \\
\frac{1}{\beta^2}\frac{\alpha'}{\alpha}\frac{\gamma'}{\gamma}&=&
+\frac{1}{\Lambda}\frac{1}{\beta^2}\left(\varphi_{*}'\right)^2+
\frac{\tilde{\omega}^2}{\alpha^2}\varphi_{*}^2-\varphi_{*}^2-
\frac{1}{2}\varphi_{*}^4+C_{*}, \\
\frac{1}{\alpha\beta}\left(\frac{\alpha'}{\beta}\right)'&=&
-\frac{1}{\Lambda}\frac{1}{\beta^2}\left(\varphi_{*}'\right)^2+
\frac{\tilde{\omega}^2}{\alpha^2}\varphi_{*}^2-\varphi_{*}^2-
\frac{1}{2}\varphi_{*}^4+C_{*},
\end{eqnarray}
where ${}'$ denotes the derivative with respect to $r_{*}$.

For the limit of large self-coupling, $\Lambda\rightarrow\infty$,
these equations will be reduced to%
\footnote{
For a finite $C_*$, the actual value of $C=m^2C_*/\Lambda$ becomes
infinitely small if the limit $\Lambda=\infty$ is taken literally.
We can however interpret the limit as an approximation of a 
{\it large} 
self-coupling, and $\Lambda$ is not simply taken as a mathematical
infinity. 
}
\begin{eqnarray}
\frac{\tilde{\omega}^2}{\alpha^2}\varphi_{*}&-&
\varphi_{*}-\varphi_{*}^3=0, \label{eq:EQ0}\\
\frac{1}{\beta\gamma}\left(\frac{\gamma'}{\beta}\right)'&=&
-\frac{\tilde{\omega}^2}{\alpha^2}\varphi_{*}^2-\varphi_{*}^2-
\frac{1}{2}\varphi_{*}^4+C_{*}, \label{eq:EQ1}\\
\frac{1}{\beta^2}\frac{\alpha'}{\alpha}\frac{\gamma'}{\gamma}&=&
+\frac{\tilde{\omega}^2}{\alpha^2}\varphi_{*}^2-\varphi_{*}^2-
\frac{1}{2}\varphi_{*}^4+C_{*}, \label{eq:EQ2}\\
\frac{1}{\alpha\beta}\left(\frac{\alpha'}{\beta}\right)'&=&
+\frac{\tilde{\omega}^2}{\alpha^2}\varphi_{*}^2-\varphi_{*}^2-
\frac{1}{2}\varphi_{*}^4+C_{*}. \label{eq:EQ3}
\end{eqnarray}


Now, let us solve the set of an algebraic equation and differential 
equations~(\ref{eq:EQ0}-\ref{eq:EQ3}).

{}From Eqs.~(\ref{eq:EQ2}) and (\ref{eq:EQ3}), we find
\begin{equation}
\left(\frac{\alpha'}{\beta\gamma}\right)'=0.
\label{eq:EQ4}
\end{equation}
To solve this, we use the residual gauge choice on $\beta$.
For simplicity, we take%
\begin{equation}
\alpha\beta\gamma=r_{*}.
\label{eq:EQ5}
\end{equation}
Then Eq.~(\ref{eq:EQ4}) is solved as
\begin{equation}
\alpha^2(r)=A\left(r_{*}^2+r_{0}^2\right),
\label{eq:EQ6}
\end{equation}
where $A$ and $r_0$ are integration constants.
Since the constant scale factor of $\alpha$ can be absorbed
in the rescaling of $t$, we can fix the value $A=1$.
Then, for the later use, we notice that
\begin{equation}
\frac{d\alpha^2}{dr_{*}^2}=1.
\label{eq:EQ7}
\end{equation}

{}From Eqs.~(\ref{eq:EQ0}), (\ref{eq:EQ2}) and (\ref{eq:EQ6}), 
we find%
\footnote{This equation~(\protect\ref{eq:EQ8})
 is consistent with the other equations 
(\protect\ref{eq:EQ0}) and (\protect\ref{eq:EQ1}).}
\begin{equation}
\frac{\gamma\gamma'}{r_{*}}=
\frac{1}{2}\varphi_{*}^4+C_{*}.
\label{eq:EQ8}
\end{equation}
Here using Eq.~(\ref{eq:EQ7}), we obtain
\begin{equation}
\frac{d\gamma^2}{d\alpha^2}=
\frac{1}{2}\varphi_{*}^4+C_{*}.
\label{eq:EQ9}
\end{equation}

We have to solve Eq.~(\ref{eq:EQ9}) by using the solution 
for $\varphi_{*}$.
Eq.~(\ref{eq:EQ0}) can be solved easily.
The interior solution for $\varphi_{*}$ is
\begin{equation}
\varphi_{*}^2=\frac{\tilde{\omega}^2}{\alpha^2}-1.
\label{eq:EQ10}
\end{equation}
This describes the configuration of the boson field
inside a boson star.
On the other hand, outside the boson star, 
\begin{equation}
\varphi_{*}^2=0.
\end{equation}

Before solving the interior and exterior solution for the metric,
we rewrite the radial line element as
\begin{equation}
\beta dr=\sqrt{\frac{\Lambda}{m^2}}\beta dr_{*}=
\sqrt{\frac{\Lambda}{m^2}}\frac{r_{*}}{\alpha\gamma}
\frac{dr_{*}}{d\gamma}d\gamma=
\sqrt{\frac{\Lambda}{m^2}}
\frac{1}{\alpha}\frac{1}{\frac{1}{2}\varphi_{*}^4+C_{*}}d\gamma.
\label{eq:EQ11}
\end{equation}
Consequently, we get
\begin{equation}
\beta^2 dr^2=
\frac{\Lambda}{m^2}
\frac{1}{\alpha^2}
\frac{1}{\left(\frac{1}{2}\varphi_{*}^4+C_{*}\right)^2}d\gamma^2.
\label{eq:EQ12}
\end{equation}

First, we solve the exterior solution.
Since $\varphi_{*}=0$ outside the boson star, 
the solution of Eq.~(\ref{eq:EQ9}) is found to be
\begin{equation}
\alpha^2=\frac{1}{C_{*}}\gamma^2-B,
\label{eq:EQ13}
\end{equation}
where $B$ is an integration constant.

The full line element is obtained, by substituting 
Eqs.~(\ref{eq:EQ5}), 
(\ref{eq:EQ12}) and (\ref{eq:EQ13}) into (\ref{eq:met}), as
\begin{equation}
ds^2=-\frac{1}{C_{*}^2}
\left(C_{*}\gamma^2-C_{*}^2B\right)dt^2+
\frac{\Lambda}{m^2}\frac{1}{C_{*}\gamma^2-C_{*}^2B}
d\gamma^2+
\frac{\gamma^2}{\gamma_{0}^2}d\theta^2.
\label{eq:EQ14}
\end{equation}
Here we must remember that $C_{*}=C\Lambda/m^2$.
We further introduce a new radial coordinate $R$ defined as
\begin{equation}
R=\sqrt{\frac{\Lambda}{m^2}}\gamma,
\label{Rg}
\end{equation}
and set $\gamma_{0}=\sqrt{m^2/\Lambda}$. Then we find
\begin{equation}
ds^2=-\frac{1}{C_{*}^2}
\left(CR^2-C_{*}^2B\right)dt^2+
\frac{1}{CR^2-C_{*}^2B}
dR^2+
R^2d\theta^2.
\label{eq:EQ15}
\end{equation}
After rescaling $\frac{1}{C_{*}}dt\rightarrow dt$,
we find the metric is precisely the same as
the well-known BTZ vacuum solution~\cite{BTZ}.
Thus we identify the BTZ mass $M_o$ {\em of the boson star}
with
\begin{equation}
8GM_o=C_{*}^2B.
\label{eq:EQ16}
\end{equation}

Here we turn to solving the interior solution.
{}From Eqs.~(\ref{eq:EQ9}) and (\ref{eq:EQ10}),
we obtain a differential equation:
\begin{equation}
\frac{d\gamma^2}{d\alpha^2}=
\frac{1}{2}
\left(\frac{\tilde{\omega}^2}{\alpha^2}-1\right)^2+C_{*}.
\label{eq:EQ17}
\end{equation}
The solution of this equation is given by
\begin{equation}
\gamma^2=-\frac{1}{2}
\frac{\tilde{\omega}^4}{\alpha^2}-
\tilde{\omega}^2\ln\frac{\alpha^2}{\tilde{\omega}^2}+
\frac{1}{2}\alpha^2+
C_{*}\left(\alpha^2+D\right),
\label{eq:EQ18}
\end{equation}
where $D$ is a constant.
The value of $D$ is determined from the condition at
the boundary of the boson star, where $\varphi_{*}=0$.
One can find the condition for a smooth connection of
the exterior and interior solutions (from Eqs.~(\ref{eq:EQ12}),
(\ref{eq:EQ13}) and (\ref{eq:EQ16})):%
\footnote{The first derivative is automatically connected
when this condition (\ref{eq:EQ19}) is satisfied.}
\begin{equation}
\left.C_{*}\gamma^2-8GM_o\right|_{\alpha^2=\tilde{\omega}^2}=
\left.\alpha^2
\left(\frac{1}{2}\varphi_{*}^4+C_{*}\right)^2
\right|_{\alpha^2=\tilde{\omega}^2}.
\label{eq:EQ19}
\end{equation}
{}From Eqs.~(\ref{eq:EQ19}) and (\ref{eq:EQ18}), we get
\begin{equation}
C_{*}^2D=8GM_o.
\label{eq:EQ20}
\end{equation}

What does the metric at the center of the boson star look like?
As seen in Eq.~(\ref{eq:EQ15}), the origin is located at
$R=0$, or $\gamma=0$.
Therefore $\alpha_{0}=\alpha(\gamma=0)$ satisfies the following
equation:
\begin{equation}
-\frac{1}{2}
\frac{\tilde{\omega}^4}{\alpha_0^2}-
\tilde{\omega}^2\ln\frac{\alpha_0^2}{\tilde{\omega}^2}+
\frac{1}{2}\alpha_0^2+
C_{*}\alpha_0^2+\frac{1}{C_{*}}8GM_o=0.
\label{eq:EQ21}
\end{equation}
Thus the metric at the center of the boson star is
written by using $\alpha_0$ as
\begin{equation}
ds^2(R\approx 0)\approx -\alpha_0^2dt^2+
\frac{1}{\alpha_{0}^2}
\frac{1}{\left(\frac{1}{2}\varphi_{*0}^4+C_{*}\right)^2}
dR^2+
R^2d\theta^2,
\label{eq:EQ22}
\end{equation}
where $\varphi_{*0}^2=\frac{\tilde{\omega}^2}{\alpha_0^2}-1$
is the value of $\varphi_{*}^2$ at the center of the star.
We can give an arbitrary value for $\varphi_{*0}^2$
by choosing the value of $\tilde{\omega}$.
Thus we can take $\varphi_{*0}^2$ as a free parameter
which characterizes a solution for a boson star, instead of
$\tilde{\omega}$.


In higher dimension more than three,
we can fix the natural value of the metric component at the center.
In three dimensions, we cannot take a definite choice,
because there is no {\em local} definition of mass;
an additional constant is allowed because it does not lead to a 
curvature singularity directly, unlike the  higher-dimensional cases.
This is also the reason why an arbitrary constant remains in the study
of Cruz and Zanelli~\cite{CZ} on relativistic stars in $(2+1)$ 
dimensions.

Nevertheless, we can choose a plausible choice
of the metric at the center,%
\footnote{
The conical sigularity at the origin is not {\em a priori} governed
by the {\em differential} equations because only the boundary 
condition is concerned. This is even true for a vacuum case.}
in order to give the definite properties of
the boson star; if one wishes to know the general case,
one need only loose the constraint in the following analysis.

We choose the following condition:
\begin{equation}
\alpha_{0}^2
\left(\frac{1}{2}\varphi_{*0}^4+C_{*}\right)^2=1.
\label{eq:EQ23}
\end{equation}
According to this choice,
the spatial metric at the origin looks like the one with no
conical singularity.
The condition~(\ref{eq:EQ23}) determines the value of 
$\alpha_0$, as a function of $\varphi_{*0}$.

This choice for the boundary condition makes the mass of the scalar 
boson
definite.  In other words, if we take other values, the parameter $m$
does not represent the mass of the excitation of the boson field
$\varphi$ near the origin; the value must be rescaled.
In the present case, since there is no asymptotically flat spacetime
region, the identification of mass parameter in the most interior 
region is reasonable.


We obtain the relation between the BTZ mass of the boson star 
and $\varphi_{*0}$ from Eqs.~(\ref{eq:EQ21}) and (\ref{eq:EQ23}):
\begin{equation}
\frac{8GM_o}{C_{*}}=
\frac{1}{\left(\frac{1}{2}\varphi_{*0}^4+C_{*}\right)^2}
\left[\frac{1}{2}\varphi_{*0}^4+\varphi_{*0}^2-
\left(\varphi_{*0}^2+1\right)\ln\left(\varphi_{*0}^2+1\right)-
C_{*}\right].
\label{eq:EQ24}
\end{equation}

The value of BTZ mass $M_o$ is negative for small $\varphi_{*0}$.
In particular, as $\varphi_{*0}$ goes to zero, $M_o$ approaches 
$-1/(8G)$.
One can take
\begin{equation}
M_i=M_o+\frac{1}{8G},
\label{eq:EQ25}
\end{equation}
as a definition of the intrinsic mass $M_i$, we call here.
When $\varphi_{*0}$ vanishes, $M_i$ becomes zero,
and spacetime becomes a usual anti-de Sitter spacetime 
everywhere, i.e.,
\begin{equation}
ds^2=-\left(1+CR^2\right)dt^2+\frac{dR^2}{1+CR^2}+
R^2d\theta^2,
\label{eq:EQ26}
\end{equation}
where the temporal coordinate has been appropriately
normalized. Then we have so-called AdS vacuum.%
\footnote{Note that, in the presence of matter,
the limit of ``no matter'' cannot yields
so-called black hole vacuum, which is given by
BTZ geometry in the limit $M_o=0$~\cite{CZ}.}

\begin{figure}
\centering
\epsfbox{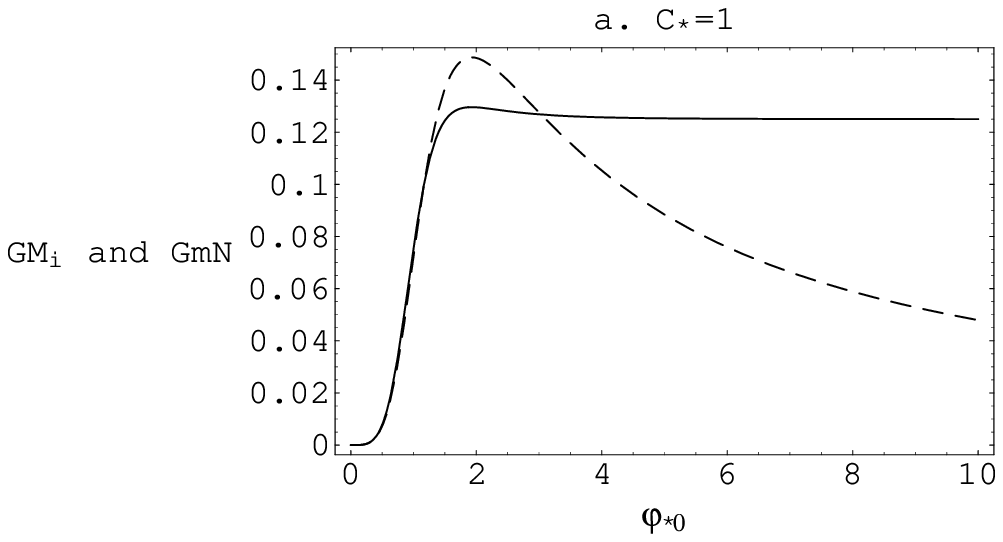}
\epsfbox{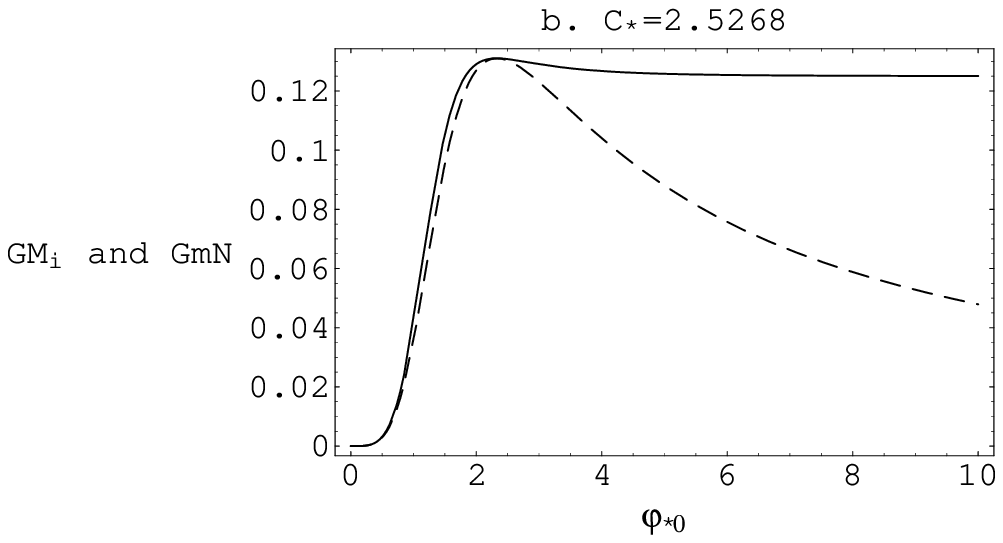}
\epsfbox{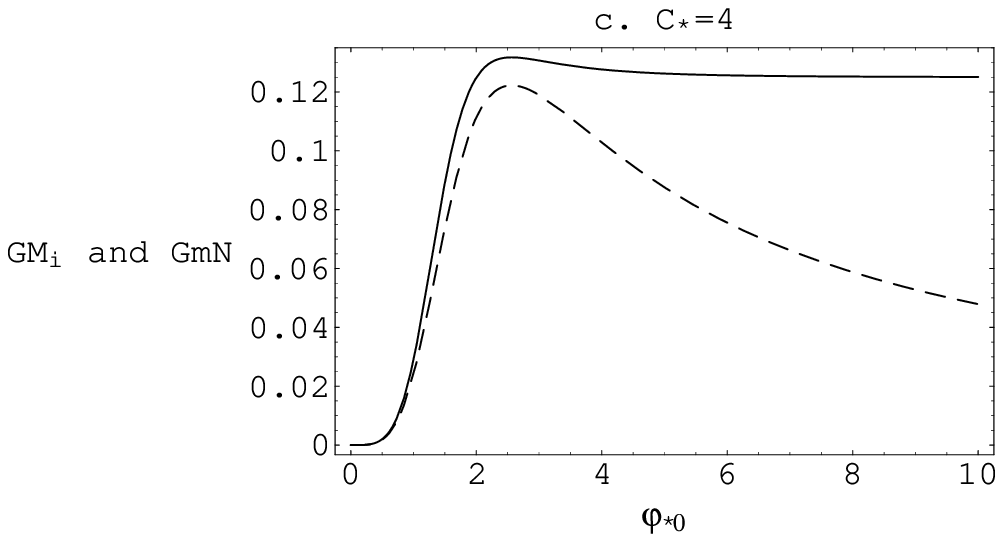}
\caption{Boson star mass $M_i$ 
in units of $G^{-1}$ (solid line) and particle number $N$ 
in units of $(Gm)^{-1}$ (broken line)
as functions of $\varphi_{*0}$ for different values of $C_{*}$:
(a) $C_{*}=1$. (b) $C_{*}=2.5268$. (c) $C_{*}=4$.}
\label{fig1}
\end{figure}

The total particle number $N$ is given by
\begin{equation}
N=\int d^2x \sqrt{-g}\left|g^{tt}\right|i\left(
\varphi\partial_t\varphi^{*}-\varphi^{*}\partial_t\varphi\right).
\end{equation}
Therefore we can calculate the particle number of the boson star
by making use of the above solution. It is given by
\begin{eqnarray}
N&=&
2\pi\int\frac{2m\tilde{\omega}}{8\pi G\Lambda}\varphi_{*}^2
\frac{1}{\alpha^2}\frac{\alpha\beta\gamma}{\gamma_0}dr \nonumber \\
&=&
\frac{m\tilde{\omega}}{2G\Lambda\gamma_0}
\int\left(\frac{\tilde{\omega}^2}{\alpha^2}-1\right)
\frac{1}{\alpha^2}r_{*}\sqrt{\frac{\Lambda}{m^2}}dr_{*} \nonumber \\
&=&
\frac{m\tilde{\omega}}{4G\Lambda\gamma_0}\sqrt{\frac{\Lambda}{m^2}}
\int_{\alpha_0^2}^{\tilde{\omega}^2}
\left(\frac{\tilde{\omega}^2}{\alpha^2}-1\right)
\frac{1}{\alpha^2}d\alpha^2 \nonumber \\
&=&
\frac{m}{4G\Lambda\gamma_0}\sqrt{\frac{\Lambda}{m^2}}
\alpha_0
\frac{\tilde{\omega}}{\alpha_0}
\left(\frac{\tilde{\omega}^2}{\alpha_0^2}-1-
\ln\frac{\tilde{\omega}^2}{\alpha_0^2}\right) \nonumber \\
&=&
\frac{1}{4Gm}\frac{\sqrt{\varphi_{*0}^2+1}
\left[\varphi_{*0}^2-
\ln\left(\varphi_{*0}^2+1\right)\right]}%
{\frac{1}{2}\varphi_{*0}^4+C_{*}}.
\end{eqnarray}
The intrinsic mass and particle number of the boson star 
as functions of $\varphi_{*0}$ are plotted in
FIG.~\ref{fig1}.

For small $\varphi_{*0}$, the mass and the partcle number
are given approximately by
\begin{equation}
M_i\approx mN\approx \frac{\varphi_{*0}^4}{8GC_{*}}.
\end{equation}

Both maxima of the mass and particle number as
functions of $\varphi_{*0}$
are located at $\varphi_{*0}=\varphi_{*0m}$.
$\varphi_{*0m}$ satisfies the following equation:
\begin{equation}
6C_{*}\varphi_{*0m}^2-4\varphi_{*0m}^4-\varphi_{*0m}^6+ 
\left(-2C_{*} + 4\varphi_{*0m}^2 + 3\varphi_{*0m}^4\right)
\ln\left(1+\varphi_{*0m}^2\right)=0.
\label{eq:pm}
\end{equation}
Unfortunately, the solution of this equation cannot be expressed by
the usual mathematical functions. For small $C_{*}$, we can solve 
the equation~(\ref{eq:pm}) approximately and get 
$\varphi_{*0m}\approx (24C_{*})^{1/6}$.

\begin{figure}
\centering
\epsfbox{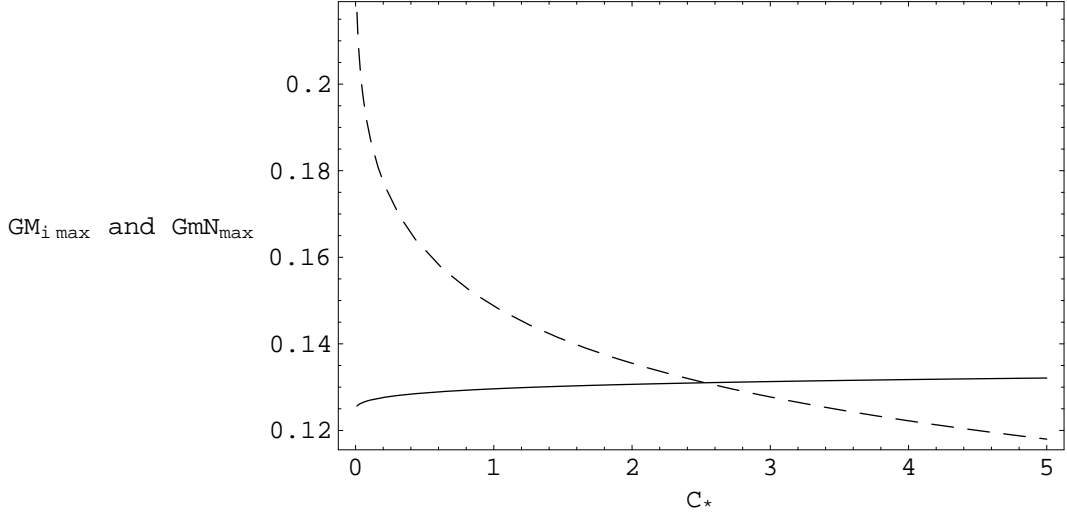}
\caption{The value of maximum intrinsic mass $M_{i~max}$
in units of $G^{-1}$ (solid line) and
The value of maximum particle number $N_{max}$
in units of $(Gm)^{-1}$ (broken line)
as functions of $C{*}$.}
\label{fig2}
\end{figure}

The maximum values of $M_i$ and $N$, which are denoted by
$M_{i~max}$ and $N_{max}$ respectively,
is shown in FIG.~\ref{fig2} as functions of $C_{*}$.
It is worth noting that the value of maximum BTZ mass 
$(M_{o~max}=M_{i~max}-\frac{1}{8G})$ is positive 
for all finite values of $C_{*}$. In particular, in the limit
$C_{*}=0$, the value of $M_{o~max}$ vanishes. The behavior of 
$M_{o~max}$ for small $C_{*}$ is given by $M_{o~max}\approx
C_{*}^{2/3}/(32\sqrt[3]{3}G)$.
The value of $N_{max}$ in the limit $C_{*}=0$ is $1/(4Gm)$.

The binding energy $E_{b~max}=M_{i~max}-mN_{max}$ 
for the maximum case 
is negative for sufficiently small $C_{*}$, as seen 
from FIG.~\ref{fig2}.
This implies that the boson star solution for small $C_{*}$ is
energetically stable. When the value of $C_{*}$ is larger than a
critical value $C_{*~crit}$, $E_{b~max}$ turns to be positive.
We find that the critical value $C_{*~crit}$ is
\begin{equation}
C_{*~crit}=
\frac{16}{243}
\frac{1598+434\sqrt{13}-\left(1242+351\sqrt{13}\right)\ln
\frac{4+\sqrt{13}}{3}}%
{10+4\sqrt{13}-3\ln\frac{4+\sqrt{13}}{3}}\approx 2.5268.
\end{equation}

\begin{figure}
\centering
\epsfbox{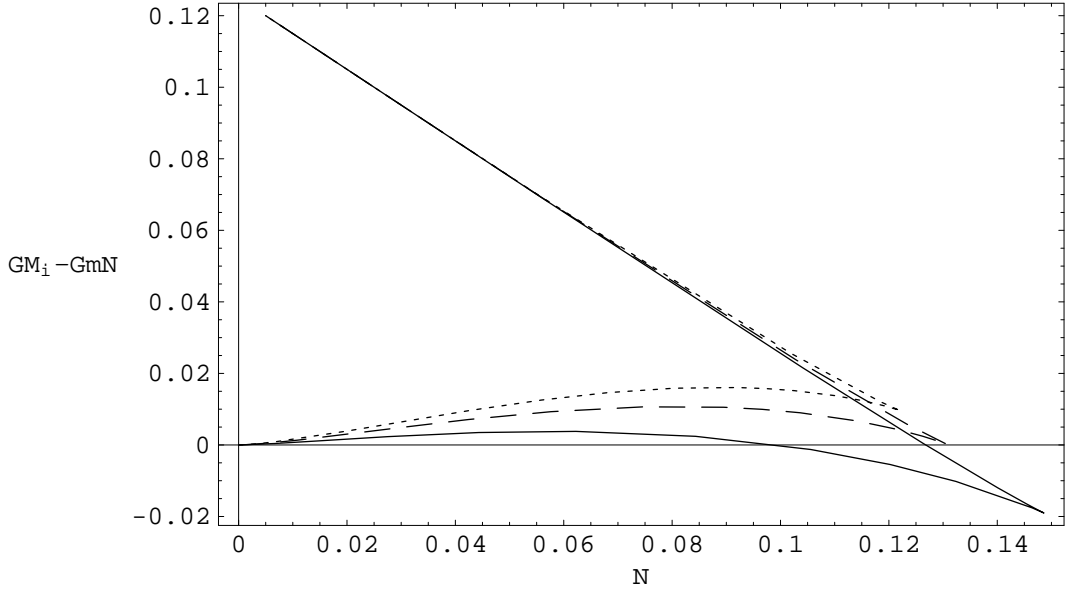}
\caption{Binding energy $E_b$ (in units of $G^{-1}$)
as a function of $N$ (in units of $(Gm)^{-1}$)
for $C_{*}=1$ (solid line), $C_{*~crit}(\approx
2.5268)$  (broken line), and $4$ (dotted line).}
\label{fig3}
\end{figure}

To see the meaning of the critical value,
we plot the dependence of binding energy $E_b=M_{i}-mN$
on the particle number $N$ for a fixed $C_{*}$.
FIG.~\ref{fig3} shows the so-called bifurcation diagrams~\cite{KMS}.
The cuspoidal point corresponds to the boson star
with the mass $M_{i~max}$ and the particle number $N_{max}$.
This point represents an absolute stable solution~\cite{KMS}.
When $C_{*}>C_{*~crit}$, however, the cuspoidal point
is not a global minimum of the binding energy.
In this case, the boson star is not stable
because the binding energy is positive for all finite $N$.%
\footnote{Notice also that the point corresponding to $N=0$ and 
$E_b=0$ is always a minimum point, because $E_b\approx 
+\varphi_{*0}^6/(24GC_{*})$ for small $\varphi_{*0}$.}

\begin{figure}
\centering
\epsfbox{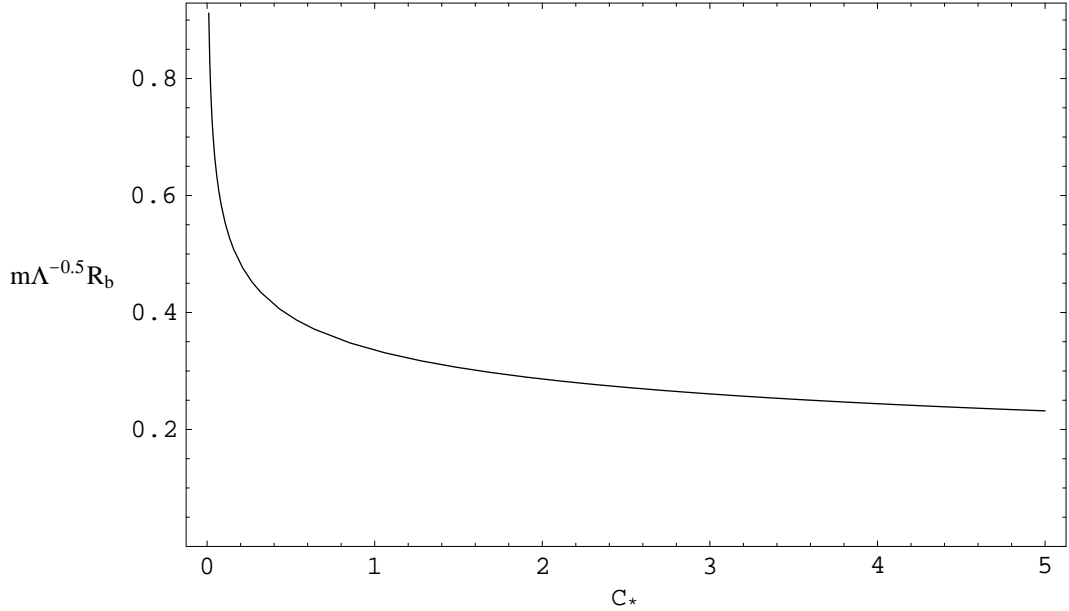}
\caption{The radius of the boson star $R_b$
(in units of $\sqrt{\Lambda/m^2}$)
with maximum mass $M_{i~max}$ as a function of $C_{*}$.}
\label{fig4}
\end{figure}

Lastly, we discuss the size of the boson star.
We define the radius of the boson star by the radius of the boundary 
of the interior and exterior solutions, where $\varphi_{*}=0$.
The radius of the boson star $R_b$ is derived from 
Eqs.~(\ref{eq:EQ18}) and (\ref{eq:EQ20}) with the 
definition~(\ref{Rg}). We find
\begin{eqnarray}
R_b^2&=&\frac{\Lambda}{m^2}\left[
\frac{C_{*}\left(\varphi_{*0}^2+1\right)}%
{\left(\frac{1}{2}\varphi_{*0}^4+C_{*}\right)^2}+
\frac{8GM_o}{C_{*}}\right] \nonumber \\
&=&\frac{\Lambda}{m^2}
\frac{1}{\left(\frac{1}{2}\varphi_{*0}^4+C_{*}\right)^2}
\left[\frac{1}{2}\varphi_{*0}^4+\varphi_{*0}^2-
\left(\varphi_{*0}^2+1\right)\ln\left(\varphi_{*0}^2+1\right)+
C_{*}\varphi_{*0}^2\right].
\label{eq:Rb}
\end{eqnarray}
The radius of the boson star with maximum mass
as a function of $C_{*}$ is plotted in FIG.~\ref{fig4}.
For small $C_{*}$, $R_b$ of the boson star with maximum mass
behaves as
$R_b\approx\sqrt{\frac{\Lambda}{m^2}}\frac{1}{3^{2/3}C_{*}^{1/6}}$.

\begin{figure}
\centering
\epsfbox{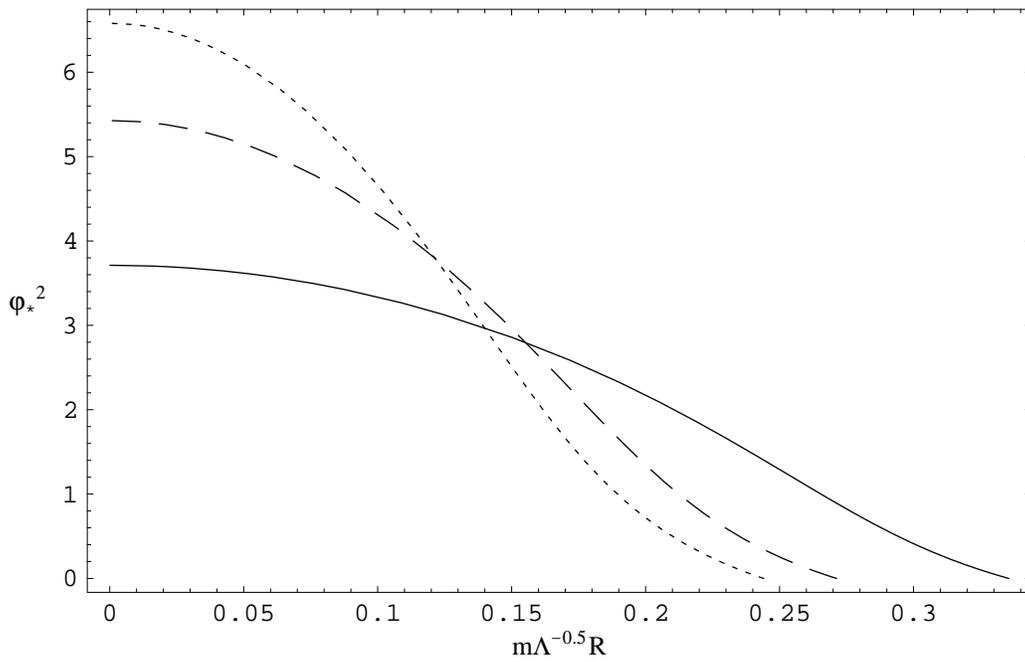}
\caption{The behavior of the scalar field $\varphi_{*}^2$ as
a function of $R$ (in units of $\sqrt{\Lambda/m^2}$)
 for a boson star with maximum mass $M_{i~max}$
for $C_{*}=1$ (solid line), $C_{*~crit}(\approx 2.5268)$ 
(broken line), and $4$ (dotted line).}
\label{fig5}
\end{figure}

The profile of the boson field can also be solved in the same way.
That is given by
\begin{equation}
R^2=\frac{\Lambda}{m^2}
\frac{1}{\left(\frac{1}{2}\varphi_{*0}^4+C_{*}\right)^2}
\left\{\left[\frac{1}{2}\left(\varphi_{*0}^2+1\right)+
\frac{\frac{1}{2}+C_{*}}{\varphi_{*}^2+1}\right]
\left(\varphi_{*0}^2-\varphi_{*}^2\right)+
\left(\varphi_{*0}^2+1\right)\ln
\frac{\varphi_{*}^2+1}{\varphi_{*0}^2+1}
\right\}.
\label{eq:Rp}
\end{equation}
Of course $R_b$ (\ref{eq:Rb}) is also obtained from this equation
by setting $\varphi_{*}=0$.
One can see from (\ref{eq:Rp}) that $\alpha$ has no singular behavior,
because $\alpha^2=\tilde{\omega}^2/\left(\varphi_{*}^2+1\right)$.
The profiles of the boson field $\varphi_{*}^2$ for specific values
of $C_{*}$ are shown in FIG.~\ref{fig5}.


To summarize, we have obtained an exact solution describing a boson 
star with very large self-coupling constant in $(2+1)$ dimensions.
There is a critical value for $C_{*}=(\lambda/(8\pi Gm^4))C$,
$C_{*~crit}\approx 2.5268$. For $C_{*}>C_{*~crit}$ the 
binding energy turns to be positive, the boson star configuration
is not energetically favorable.

Explicit study of stability of the boson star under linear
perturbations and pulsations\cite{BS,LM}
will be discussed in future publications.
The analysis of the boson star with an arbitrary, 
{\em actually finite} value
of the self-coupling is of much interest and will be necessary.

The rotation of the boson star%
\footnote{Properties of a spinning boson star with large 
self-interaction in $(3+1)$ dimensions is studied 
by Ryan~\cite{Ryan}.}
is easily incorporated 
in the $(2+1)$ dimensional model.
The rigorous result on a nonrotating boson star in this paper
is a basis of the study of the rotatinge case.

More general cases including such as a $|\varphi|^6$ coupling
may exhibit more complicated results,but the analysis of them
can be carried out in the same manner as in this paper.
We expect that our analysis can also be easily extended to
the solutions for fermion-boson stars in three dimensions.
It will be the subject to be studied elsewhere.

\section*{Acknowledgement}
The author would like to thank to the referee for comments.


\newpage

\end{document}